\documentclass[useAMS,usenatbib]{mn2e}

\usepackage{epsfig}
\usepackage{myaasmacros}

\title[The influence of gas on the structure of merger
  remnants]{The influence of gas on the structure of merger
  remnants} 
\author[Thorsten Naab, Roland Jesseit \& Andreas Burkert]{Thorsten Naab$^{1,2}$\thanks{E-mail:
naab@usm.lmu.de}
, Roland Jesseit$^{1}$ and Andreas Burkert$^{1}$
\\
$^{1}$Universit\"ats Sternwarte M\"unchen, Scheinerstr.1, D-81679 M\"unchen, Germany \\
$^{2}$Institute of Astronomy, Madingley Road, Cambridge, CB3 0HA, UK}

\begin{document}

\date{Accepted ???. Received ??? in original form ???}

\pagerange{\pageref{firstpage}--\pageref{lastpage}} \pubyear{2002}

\maketitle

\label{firstpage}

\begin{abstract}
We present a large set of merger simulations of early-type disc galaxies with mass ratios of 
1:1 and 3:1 and 10\% of the total disc mass in gas. The internal orbital structure and
the kinematic and photometric properties of the remnants are analysed in detail
and compared to pure stellar mergers. In contrast to the collisionless case 
equal-mass mergers with gas do not result in very boxy remnants which is caused 
by the suppression of box orbits and the change of the projected shape of 
minor-axis tube orbits in the more axisymmetric remnants. The isophotal shape of 3:1 
remnants and the global kinematic properties of 1:1 and 3:1 remnants are only 
weakly affected by the presence of gas. 1:1 remnants are slowly rotating 
whereas 3:1 remnants are fast rotating and discy. The shape of the stellar LOSVD
is strongly influenced by gas. Within the effective radius the LOSVDs of collisionless 
remnants have broad leading wings while their gaseous counterparts show steep leading wings,
more consistent with observations of elliptical galaxies. We show that this change 
is also caused by the suppressed populating of box orbits 
and it is amplified by the formation of extended gas discs in the merger remnants 
which might eventually turn into stars. If elliptical galaxies have 
formed from mergers our results indicate that massive, slowly rotating boxy 
elliptical galaxies can not have formed from dissipative mergers of discs. Pure stellar 
(dry) mergers are the more likely candidates. On the other hand lower mass, fast rotating 
and discy ellipticals can have formed from dissipative (wet) mergers of early-type discs. 
So far, only unequal-mass disc mergers with gas can successfully explain their observed 
substructure. This is consistent with the revised morphological classification 
scheme of increasing importance of gas dissipation when moving from boxy ellipticals 
to discy ellipticals and then to spiral galaxies, proposed by Kormendy \& Bender.
\end{abstract}


\section{Introduction}
Inspired by the work of \citet{1972ApJ...178..623T} a large number 
of simulations have been performed 
to investigate whether giant elliptical galaxies can be formed by  
binary mergers of disc galaxies. If this formation is correct the properties 
of the merger remnants would have to be in agreement with 
all observed properties of elliptical galaxies: their global
kinematic and photometric properties as well as their fine-structure. 
Violent relaxation \citep{1967MNRAS.136..101L,2005MNRAS.362..252A} of 
the stellar and dark matter components is the dominant dynamical process 
in the rapidly varying potential of an ongoing merger. However, it does not 
completely erase the information about the progenitor galaxies. 
Therefore we should  observe signposts of the formation process in the remnants, 
e.g. the morphology of the progenitor galaxies or the encounter geometry, 
both in the dynamics and the photometric properties. Elliptical galaxies, 
if they have formed this way, should show similar properties, providing information 
about the properties of their progenitors.  

Collisionless simulations of equal mass mergers of disc galaxies 
have been studied in great detail by several authors 
(e.g. \citealp{1981MNRAS.197..179G,1992ApJ...393..484B, 
1992ApJ...400..460H,2003ApJ...597..893N}, hereafter 
NB03,\citealp{2005MNRAS.357..753G},  \citealp{2005MNRAS.360.1185J} hereafter JNB05, 
\citealp{2006MNRAS.369..625N}, \citealp{2005MNRAS.357..753G}).  
The global properties of the remnants are in agreement with observations of giant 
elliptical galaxies in the intermediate mass range, e.g. equal mass
remnants are slowly rotating, anisotropic, have boxy or discy
isophotes. Unequal mass mergers are more isotropic and have
discy isophotes (NB03). Merger remnants in general have phase space densities and 
surface density profiles that resemble observed ellipticals if bulges are 
added to the progenitor discs \citep{1993ApJ...416..415H,2006MNRAS.369..625N}. 

As soon as the kinematics of simulated merger remnants was investigated 
in more detail, an interesting disagreement with observed elliptical galaxies 
was revealed. The line-of-sight velocity distributions (LOSVD) within 
the effective radius of merger remnants in general show small asymmetric 
deviations from Gaussian shape. 
They tend to have a steep trailing wing \citep{2000MNRAS.316..315B,2001ApJ...555L..91N}. 
In contrast, most observed rotating ellipticals clearly show a 
steep leading wing in their LOSVDs \citep{1994MNRAS.269..785B}. 
Theoretically, axisymmetric, rotating one-component systems in fact show such a 
behaviour \citep{1994MNRAS.268.1019D}. This would indicate that ellipticals are very simple 
one component systems that did not form by mergers, It is, however, unlikely that 
elliptical galaxies are such simple systems (see e.g. \citealp{2004MNRAS.352..721E}). 
An alternative explanation, based on photometric and kinematical 
observations, is that rotating ellipticals contain embedded 
large scale stellar discs (e.g. \citealp{1990ApJ...362...52R,1998A&AS..131..265S,
1999ApJ...513L..25R}). A superposition of two distinct components, 
e.g. a hot spheroidal bulge and a rotationally supported cold disc, 
can also result in a steep leading wing of the LOSVD \citep{1994MNRAS.269..785B}. 
\citet{2001ApJ...555L..91N} showed that embedding an exponential 
stellar disc artificially in collisionless merger remnants 
would change the asymmetries in the LOSVDs, leading to a good agreement 
with observations. Those discs need a scale length similar to the effective 
radius of the bulge and 10\% -20\% of its mass.  

Such an extended disc naturally forms in merger remnants from high angular momentum gas 
located in the outer regions of the progenitors after it has been expelled 
along the tidal arms \citep{2001ASPC..230..451N,2002MNRAS.333..481B,
2005ApJ...622L...9S,2005astro.ph.10821W}. 
Gas initially located at smaller radii shocks and falls to the center of the
remnant probably causing a starburst and/or feeding a super-massive black
hole \citep{1996ApJ...464..641M,2005ApJ...620L..79S}. This gas does not 
contribute to the extended disc.  
It has been shown by \citet{1998ApJ...502L.133B} using simulations including gas and 
star formation that rotating lenticular galaxies with a disc like component can form 
from unequal-mass disc mergers.  Furthermore, \citet{2005ApJ...622L...9S} has 
demonstrated with a simulation including star formation and feedback that a 
remnant with a dominant stellar 
discs can form self-consistently in an equal-mass merger of two gaseous discs. 
However, those initial conditions might be more typical for high redshift discs and for 
such an extreme case starformation can not be neglected. 

In this paper we present simulations 
of disc galaxy mergers with small gas fractions of 10\% which are more similar 
to low redshift discs. Our aim is to understand the influence of a small 
dissipative component (which will alway exist, even if star formation is considered) 
on the global structure of mergers remnants. This will help to constrain the 
importance of additional physical processes like star formation, black-hole accretion 
and feedback (e.g. \citealp{2005astro.ph..3201C,2005ApJ...622L...9S,2006ApJ...641...21R}). 
It has been shown by \citet{1996ApJ...471..115B} that the presence of gas 
influences the stellar component of the merging galaxies. 
Gas that accumulates at the center of merger remnants changes the 
central potential and can increase the velocity dispersion of the stars.
\citet{2006ApJ...641...21R} have demonstrated this effect in merger 
simulations including star formation. In addition, a steeper central 
potential well results in a more axisymmetric central shape of the remnants 
\citep{1998giis.conf..275B}. 
At the same time the fraction of stars on box orbits 
is significantly reduced and tubes become the dominant orbit family 
\citep{1996ApJ...471..115B}. The most reasonable explanation for this 
behaviour is that systems with steep cusps in their potential can not 
sustain a large population of box orbits \citep{1985MNRAS.216..467G,1993ApJ...409..563S,1996ApJ...460..136M,1998ApJ...506..686V,1998giis.conf..275B}. 

In this paper we present a new and more detailed analysis. With respect to 
resolution, statistical  
completeness and comparison to observations our study goes beyond previous 
investigations of similar simulations (e.g. \citealp{1983MNRAS.205.1009N,1992ApJ...393..484B,
1996ApJ...471..115B,2000MNRAS.316..315B}).  We show quantitatively how 
a dissipative component changes the LOSVDs and the photometric 
properties of the merger remnants and explain in detail 
how these changes are connected to the changes of the orbital content of 
the remnants (see JNB05 for collisionless remnants). In particular, 
we demonstrate that a dissipative component can explain  
the origin of the observed asymmetries in the LOSVDs of elliptical galaxies. This paper 
is accompanied by a detailed two-dimensional kinematic analysis of the remnants presented
here \citep{2006astro.ph..6144J}.

The paper is organised as follows. Section \ref{MODELS} presents
the details of the simulations. A short overview of observable
properties of ongoing mergers is given in Section \ref{COMMENTS}. The
intrinsic and projected shapes of the remnants and their orbital
content are discussed in Section \ref{SHAPES}. In Section \ref{PHOTO} we
show the projected photometric and kinematic properties of the
remnants followed by a detailed investigation of the LOSVDs in Section
\ref{LOSVD}. A summary and conclusions follow in Section \ref{SUMMARY}.  

\section{The merger models}
\label{MODELS}

\begin{figure}
\begin{center}
  \epsfig{file=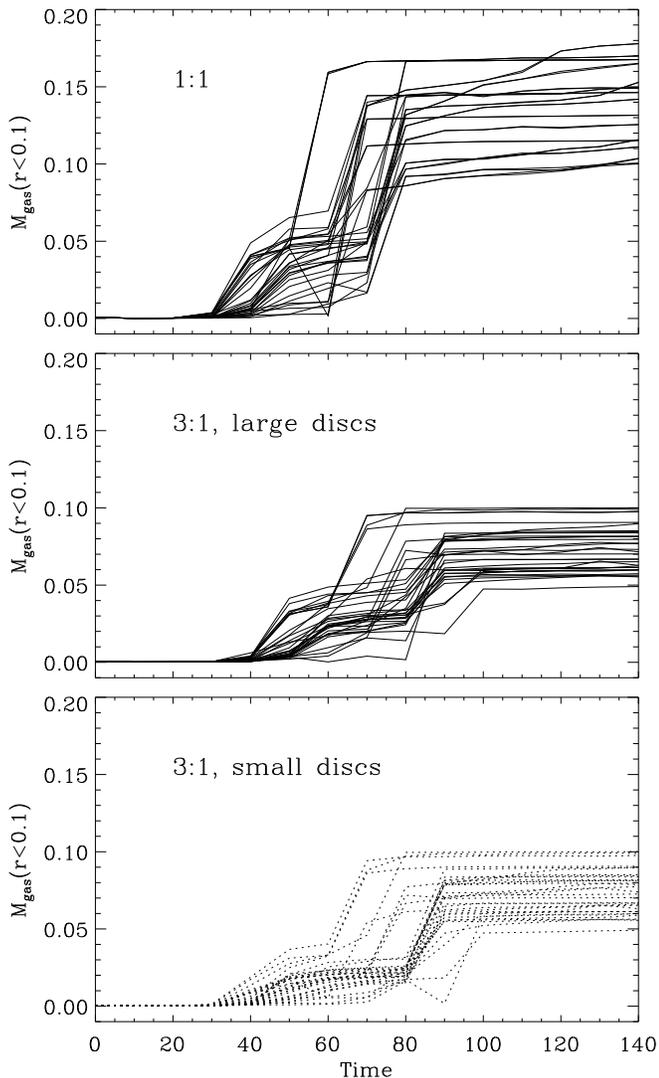, width=0.5\textwidth}
  \caption{Time evolution of the gas mass inside a radius of $r = 0.1$ 
of the galaxies for 1:1 mergers (top panel) and the large (middle panel) 
and small (bottom panel) progenitor galaxies of 3:1 mergers. The central region 
of each galaxy has been followed separately. As the galaxies merge the enclosed masses
become identical. The total initial gas mass of the more massive discs is 
$M_{\mathrm{gas}}=0.1$ and $M_{\mathrm{gas}}=0.0333$ for the less
massive discs.  
  {\label{time_vs_mgas_all}}}
\end{center}
\end{figure}
Disc galaxies were constructed in dynamical equilibrium using the
method described by \citet{1993ApJS...86..389H}. The system of units
was: gravitational constant G=1, exponential scale length of the
larger progenitor disc $h_d=1$ (the scale height was $h_z=0.2$) and
mass of the larger disc $M_d=1$. The discs were exponential with an
additional spherical, non-rotating bulge with mass $M_b = 1/3$, a
Hernquist density profile \citep{1990ApJ...356..359H} and a scale
length $r_b=0.2h_d$, and a pseudo-isothermal halo with a mass $M_d=5.8$,
cut-off radius $r_c=10h_d$ and core radius $\gamma=1h$. The parameters
for the individual components were the same as for the collisionless
mergers presented in NB03. Scaled to
the Milky Way, one length unit is $3.5kpc$, one
velocity unit is $262km/s$ and the mass unit is $5.6\times
10^{10} M_{\odot}$. For this study we re-simulated the full set
of 1:1 and 3:1 mergers with an additional gas component in the
disc. We replaced 10\% of the stellar disc by gas with the same scale
length and an initial scale height of $h_{z,gas} = 0.1 h_z$. The gas
was represented by SPH particles adopting an isothermal equation of
state, $P = c_s^2\rho$, with a fixed sound speed of $c_s=0.039$ in
velocity units, corresponding to $c_s \approx 10km/s$ if scaled to a
Milky Way type galaxy. Assuming an isothermal equation of state
implies that additional heat created in shocks, by adiabatic
compression and feedback processes is radiated away immediately. It
also implies that substantial heating processes prevent the gas from
cooling below its effective temperature, e.g. sound speed. Recently, a
number of simulations have shown that using an isothermal equation  of
state is a reasonably good approximation to the ISM in disc galaxies
(see e.g. \citealp{2001ASPC..230..451N,2002MNRAS.333..481B,2005ApJ...620L..19L}
and references therein).  
 
We followed mergers of discs with mass ratios of $\eta =1$ and $\eta
=3$ where $\eta $ is the mass of the more massive galaxy divided by
the mass of the merger partner. The equal-mass mergers were calculated
adopting in total 440000 particles with each galaxy consisting of
20000 bulge particles, 60000 stellar disc particles, 20000 SPH
particles representing the gas component in the disc, and 120000 halo
particles.  We decided to use twice as many halo particles than disc
particles to reduce heating and instability effects in the disc
components \citep{1999ApJ...523L.133N} by encounters between halo
and disc particles. For the mergers with $\eta =3$
the parameters of the more massive galaxy were as described above. The
low-mass companion contained a fraction of $1 / \eta $ the mass and
the number of particles in each component, with a disc scale-length
(stars and gas) of $h=\sqrt{1/\eta }$, as expected from the
Tully-Fisher relation \citep{1992ApJ...387...47P}. 

The N-body/SPH simulations were performed using the hybrid N-body/SPH
treecode VINE (Wetzstein et al., in preparation)  with individual timesteps. 
The gravitational forces were
softened with a Spline kernel of $h_{\mathrm{grav}} = 0.05$. The
minimal size of the Spline kernel used for computing the SPH
properties, $h_{\mathrm{SPH}}$, was fixed to the same
value. Implicitly, this procedure suppressed gas collapse on scales
smaller than the softening scale and prevents numerical instabilities
\citep{1997MNRAS.288.1060B}. All simulation have been run on a cluster of
64 1.5GHz Sun CPUs at the Institute of Astronomy in Cambridge.   

The initial discs were run in isolation for two dynamical times to allow 
the systems to finally settle into an equilibrium state. In the merger  
the galaxies approached each other on nearly parabolic orbits with an
initial separation of 30 length units and a pericenter distance of 2
length units. A study of orbits of merging dark matter halos in
cosmological large scale simulations by \cite{2006A&A...445..403K} has
shown that a significant number of the merging halos are  indeed on
parabolic orbits with a broad distribution of pericenter distances. In this 
study we focus on the usually in simulations considered regime of
small pericenter distances. In selecting unbiased initial
parameters for the disc inclinations we followed the procedure
described by \citet{1998giis.conf..275B}. The initial orientations for
the discs were the same as in NB03, Table 1. The merger remnants were 
allowed to settle into dynamical equilibrium for approximately 30 dynamical 
time-scales after the merger was complete. Then their equilibrium state was analysed.    

\section{Comments on merger dynamics and gas inflow}
\label{COMMENTS}

The properties of the initial disc galaxies change dramatically during the 
interaction. Tidal forces lead to the formation of tidal arms and trigger
gas inflow to the center. The stellar systems are dynamically heated and finally merge
into a new type of galaxy. Most investigations have focused on the properties 
of the final merger remnants. However, recent high resolution observations of 
nearby interacting galaxies have made it possible to compare ongoing 
mergers directly to different phases of simulated interactions \citep{2006ApJ...638..745D}.

It has not been investigated in detail during which phases of the
mergers e.g. the central velocity dispersion or the effective radius 
adjusts to its final value. A
detailed knowledge of how important observables evolve helps to
constrain measurements of e.g. the mass-ratios of 
observed disc mergers \citep{2006ApJ...638..745D}, their gas accretion
history, or their starformation rates. 

We measured the observables, in particular the effective velocity dispersion 
of the progenitor discs, during the interaction
 by following every individual galaxy analysing snapshots of the
 merger in the orbital plane every 10 unit times (approximately every 10 half-mass rotation 
periods of the more massive disc). To perform the 
analysis as consistent as possible with observations
(see \citealp{2006ApJ...638..745D}) we computed the effective 
radius $r_{\mathrm{eff}}$ of every galaxy as the projected spherical half-mass 
radius of the stellar particles within 5 length scales only taking particles 
of the galaxy itself into account. Thereby we avoided unrealistically large 
values for $r_{\mathrm{eff}}$ when the galaxies overlap. 
The effective central stellar velocity dispersion $\sigma_{\mathrm{eff}}$ for each galaxy 
was then computed within $0.5 r_{\mathrm{eff}}$
taking all stellar particles into account. To follow the gas accretion  
onto the center we computed the total gas mass 
within a radius of $r= 0.1$, $M_{\mathrm{gas}} (r < 0.1)$, which is two 
times the gravitational Spline softening length of the simulations, 
$h_{\epsilon} =0.05$.
\begin{figure*}
\begin{center}
  \epsfig{file=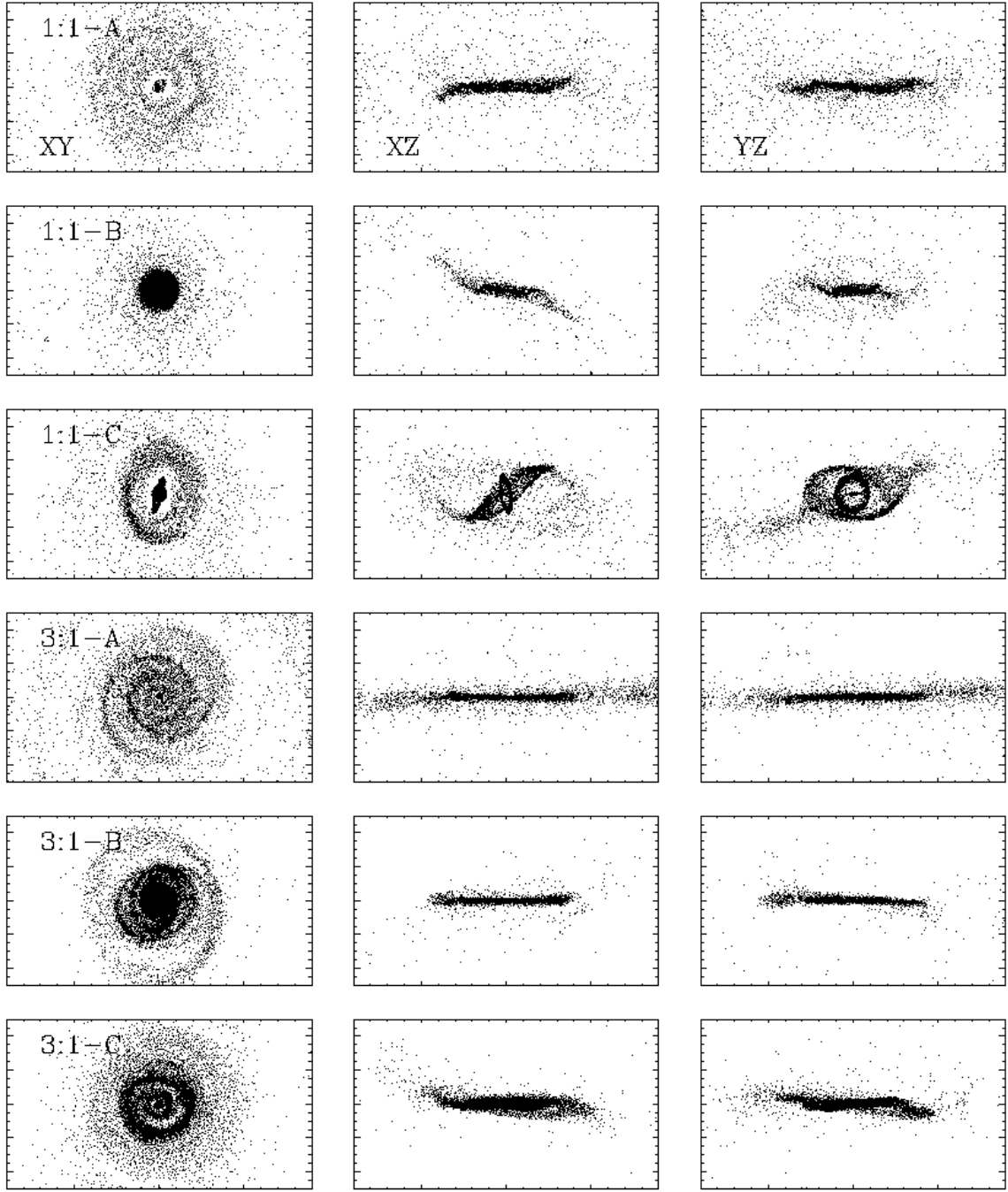, width=0.9\textwidth}
  \caption{Final gas distribution for 1:1 and 3:1 remnants with three different geometries
A, B, and C seen along the principal axes of the main stellar bodies. The box length 
is 8 unit lengths in vertical direction and 18 unit lengths in horizontal direction.  
Most remnants have embedded gas discs with a more regular disc structure for 3:1 remnants. 
  {\label{gas_all}}}
\end{center}
\end{figure*}
\begin{figure}
\begin{center}
  \epsfig{file=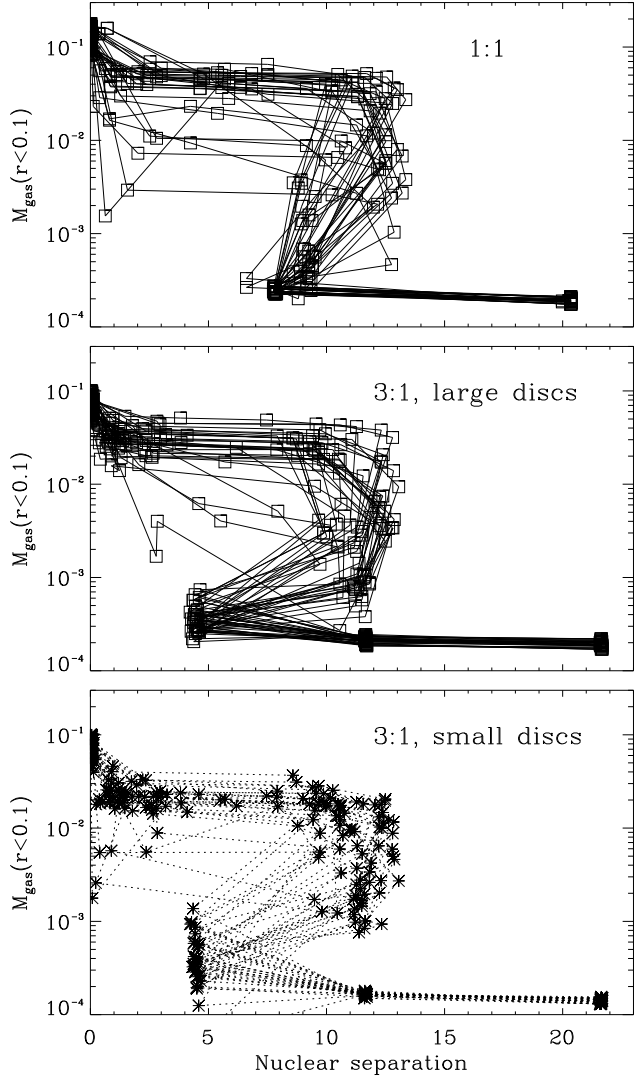, width=0.5\textwidth}
  \caption{Gas mass inside a radius of $r = 0.1$ 
of the galaxies for 1:1 mergers (top panel) and the large (middle panel) 
and small (bottom panel) progenitor galaxies of 3:1 mergers as a function 
of nuclear separation. 
  {\label{dist_vs_mgas_all}}}
\end{center}
\end{figure}
\begin{figure}
\begin{center}
  \epsfig{file=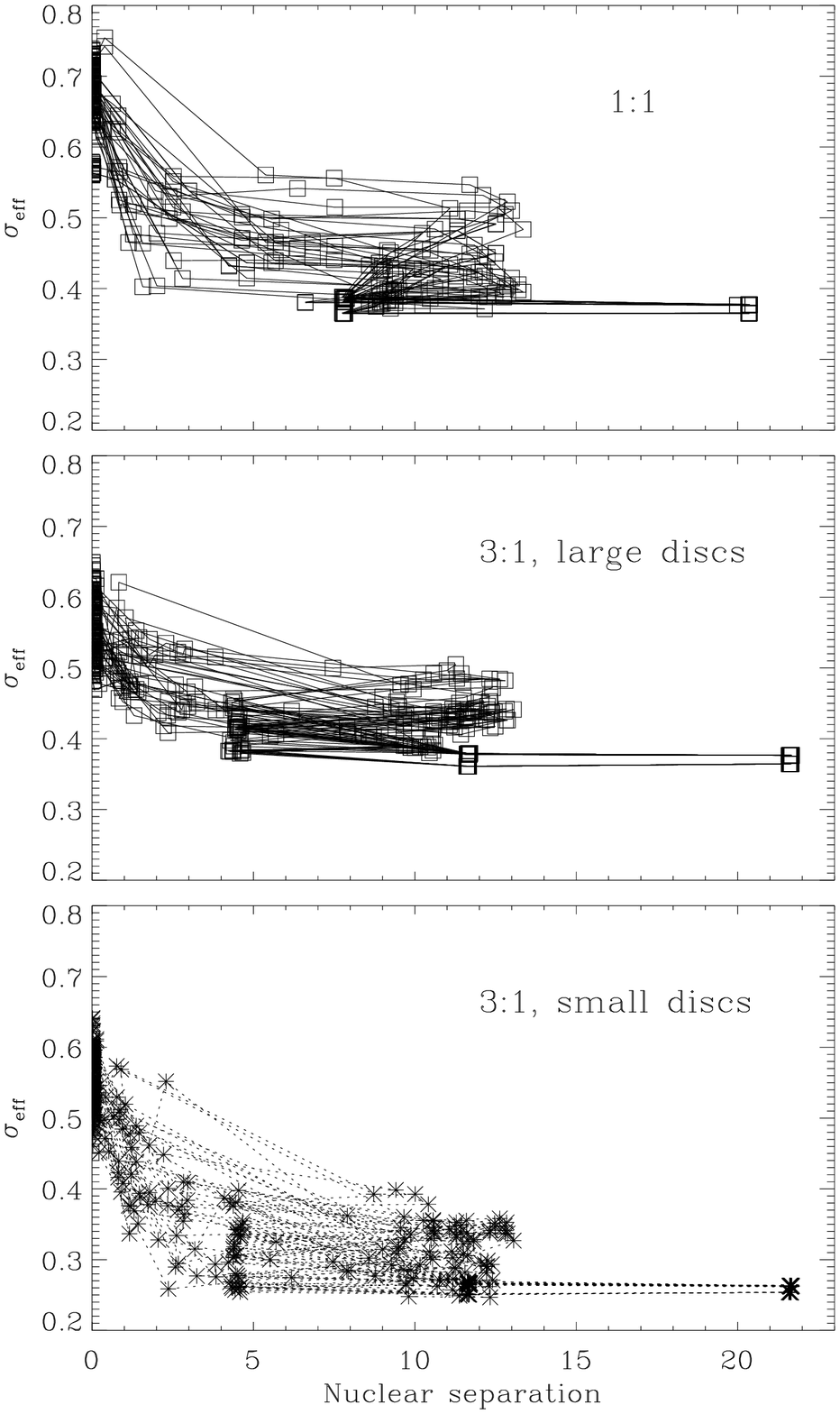, width=0.5\textwidth}
  \caption{Effective stellar velocity dispersion, $\sigma_{\mathrm{eff}}$, 
of the galaxies for 1:1 mergers (top panel) and the large (middle panel) 
and small (bottom panel) progenitor galaxies of 3:1 mergers as a function 
of nuclear separation. 
  {\label{dist_vs_sigma_all}}}
\end{center}
\end{figure}

In Fig. \ref{time_vs_mgas_all} we show the amount of gas accreted onto 
the center of each galaxy. Every line corresponds to one disc galaxy. As 
soon as the galaxies merge (at $t \approx 70$) the two lines of the progenitor galaxies 
join. In general equal mass galaxies merge faster than unequal mass galaxies. 
For 1:1 mergers every galaxy contains $M_{\mathrm{gas}}=0.1$. After the first encounter 
between 10\% and 60\% of the available gas in the progenitor discs is
funneled to the center. In some cases 
additional gas, expelled form the partner galaxy, is captured already
at early phases of the merger. The exact numbers vary 
with the encounter geometry. In general, gas transport 
is more effective if the spin and the orbital angular momentum of the disc 
is aligned \citep{2002MNRAS.333..481B}. After the galaxies have merged 
between 50\% and 85\% for 1:1 mergers 
(between 40\% and 75\% for 3:1 mergers) of the gas has accumulated at the center.

To test the effect of halo resolution on the gas inflow we have re-run two 1:1 and 3:1 
mergers with very different geometries, respectively, changing the halo properties 
in two ways: First we have changed the 
softening of the halo to $\epsilon_{\mathbf{halo}}= 0.09$ and the gas softening to 
 $\epsilon_{\mathbf{gas}}= 0.03$ to scale as the square root of the particle mass.
This guarantees that the maximum gravitational force exerted from a particle 
is independent of its mass \citep{2001MNRAS.324..273D}. The effect of this 
change on the gas properties was very small and we could not find a general 
trend. 

\begin{figure}
\begin{center}
  \epsfig{file=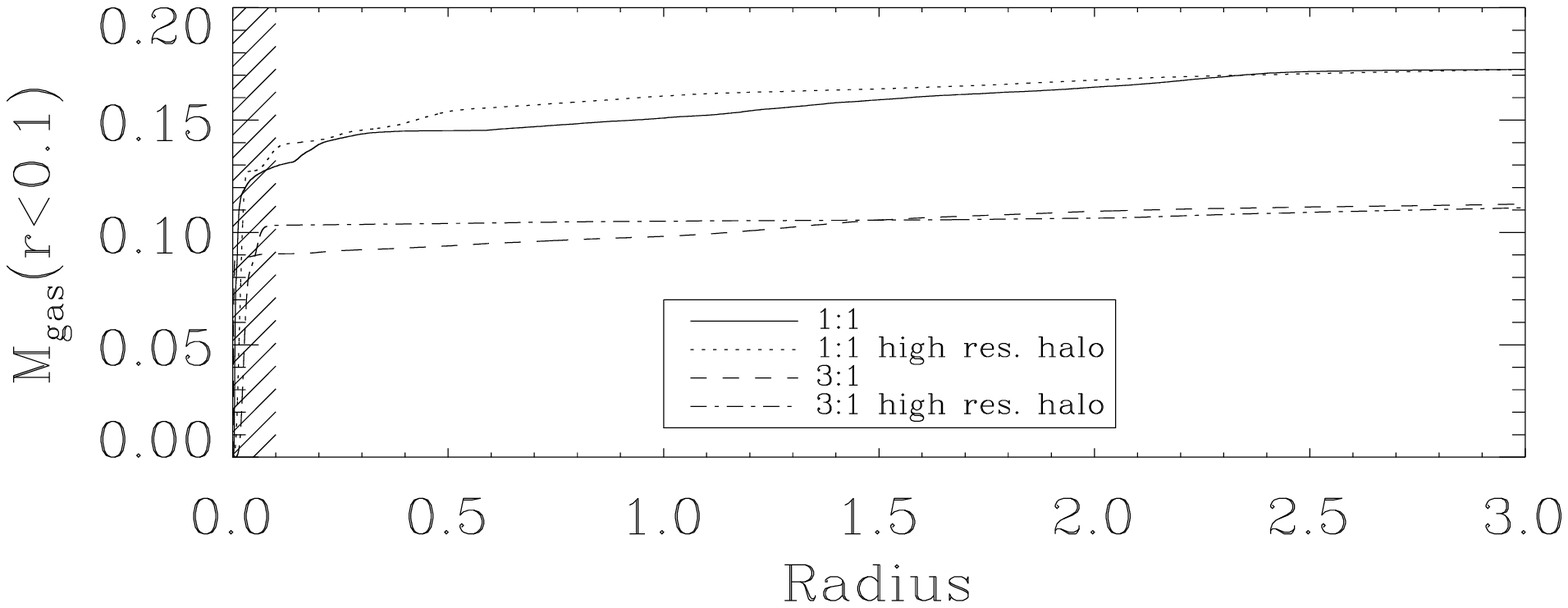, width=0.5\textwidth}
  \caption{Comparison of the cumulative gas mass versus radius for a 1:1 
(solid line) and a 3:1 merger (dashed line) and their counterparts with higher resolution in the 
halo component (dotted and dash-dotted line). The shaded area indicates the central part of 
the remnants. A better resolved halo results in a more concentrated gas distribution. The central 
gas content can increase by 15\%. 
\label{macc_comp}}
\end{center}
\end{figure}

Secondly we used the original halo softening ($\epsilon_{\mathbf{halo}}= \epsilon =0.05$) 
and increased the mass resolution of the more massive halo to 348000 particles to get the 
same individual particle masses in all collisionless components. For this set of 
simulations we found that dissipational effects are slightly enhanced. The amount of gas 
driven to the center is increased by 5\% to 15\%. In Fig. \ref{macc_comp} we show the cumulative 
gas mass distribution of one 1:1 and on 3:1 merger in comparison to the original simulation.
Most likely the higher halo resolution reduces the effect of artificial two-body heating
by encounters of halo and gas particles. 
The global effect on the gas distribution is, however, small and the stellar remnants are 
only weakly affected (see Section \ref{SHAPES}). Based on these results and to be fully consistent 
with the already performed collisionless simulations we have used a gravitational softening of 
$\epsilon =0.05$ for all components.

In Fig. \ref{gas_all} we show the final gas distribution for 1:1 and 3:1 mergers with
three different encounter geometries.  For 1:1 remnants the gas can be distributed
in warped discs or more irregular structures like polar rings. 3:1 remnants 
show a more regular disc structure. The formation of extended gas discs in 3:1 merger 
remnants has been predicted by \citet{2001ASPC..230..451N} and we
refer to \citet{2002MNRAS.333..481B} who has studied the dynamics of 
disc formation extensively. 
\begin{figure*}
\begin{center}
  \epsfig{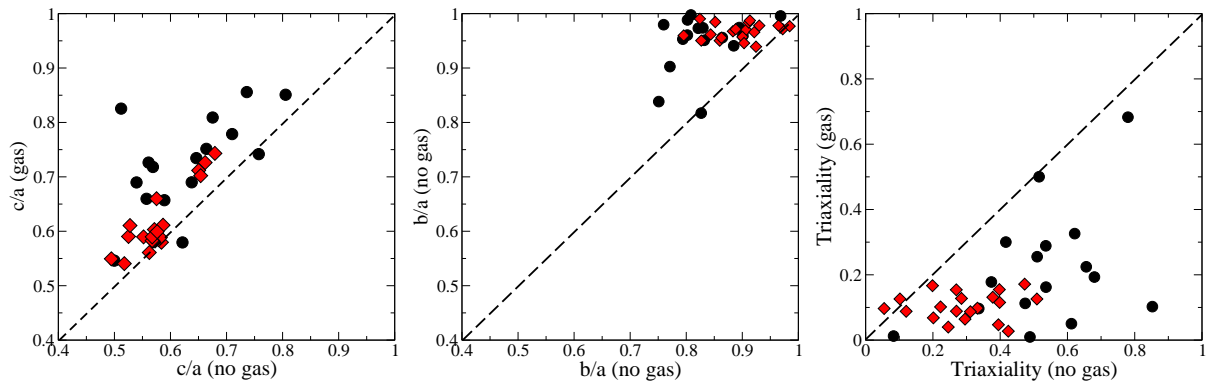}
  \caption{Axis ratios of the stellar remnants for the simulations
  with and without gas for mass ratios 1:1 (black dots) and 3:1
  (red diamonds) {\it Left panel}: ratio between the short and
  the long axis $(c/a)$. {\it Middle panel}: ratio between the intermediate and
  the long axis $(b/a)$. {\it Right panel}: triaxiality of the stellar
  remnants. Stellar remnants with a gas component are less triaxial with a similar
  maximum flattening $(c/a)$. \label{triaxiality}}
\end{center}
\end{figure*}
In Fig. \ref{dist_vs_mgas_all} we show the central gas content as a function 
of nuclear separation of the merging galaxies. The main accretion phases are 
after the first encounter and at the final merger of the galaxies 
(see e.g. \citealp{1996ApJ...464..641M}). It is 
important to note for comparisons with observations that the central gas
content differs significantly before and after the first encounter although 
the galaxies can have the same nuclear separation. This fact could be used 
as additional diagnostics to predict in which state of the merger observed 
galaxies are. The effective velocity dispersion versus nuclear separation is plotted in 
Fig. \ref{dist_vs_sigma_all}.  Similar to the gas content the dispersion rises
in two phases: after the first encounter and, to its new equilibrium value, 
in the final merger phase. In combination with the evolution of the effective 
radii these results have been used to test dynamical mass estimates of observed
nearby ULIRGs in interacting pairs \citep{2006ApJ...638..745D} which all appear 
to have mass ratios between 1:1 and 3:1. Our results also indicate that merging 
disc galaxies might already fall on the observed black-hole mass - sigma relation 
\citep{2002ApJ...574..740T} soon after their first encounter 
(Dasyra et al., submitted).

\section{Intrinsic shapes and stellar orbits of the merger remnants}
\label{SHAPES}
The intrinsic shape of a mass distribution is defined by the
ratio of its three principal axes. They were determined
by diagonalising the moment of inertia tensor of each merger
remnant. The particles were binned according to their binding energy. That
ensures that the subsets of particles follow the structure of the
remnant naturally \citep{1996ApJ...460..101W}. The triaxiality
parameter $T$ is defined as 
\begin{equation}
T=\frac{1-(b/a)^2}{1-(c/a)^2},
\end{equation}
where a, b, and c are the long, intermediate and minor-axis,
respectively. The shape of merger remnants is closely related to their
intrinsic orbital structure as shown by
JNB05. In general minor-axis
tubes are more heavily populated in oblate remnants and box orbits are
most abundant in remnants triaxial remnants with $T=0.5$ (see Fig. 5 and Fig. 6 
of JNB05). In Fig.\ref{triaxiality} we show how the presence of gas influences the
intrinsic shapes of the stellar components of the merger remnants. The
triaxiality is lowered for almost every remnant due to the
influence of gas. 3:1 remnants with gas only reach a
maximum triaxiality of $T =0.2$, while in the collisionless case they
can have triaxialities as large as $T=0.5$. Although the triaxiality of the 1:1
remnants is also lowered substantially the effect is smaller than for
3:1 remnants. The more violent merging process can to some extent counter
the dissipational influence of the gas. Gas rather makes the remnant more 
axisymmetric than more flattened (Fig.\ref{triaxiality}). The ratio between 
the short and the long axis of the triaxial body ($c/a$) is only increased 
by $\sim$ 10\%, making the remnant slightly more spherical. In contrast, for
almost all remnants $b/a$ is now close to unity, i.e. the systems are 
axisymmetric. 

\begin{figure}
\begin{center}
  \epsfig{file=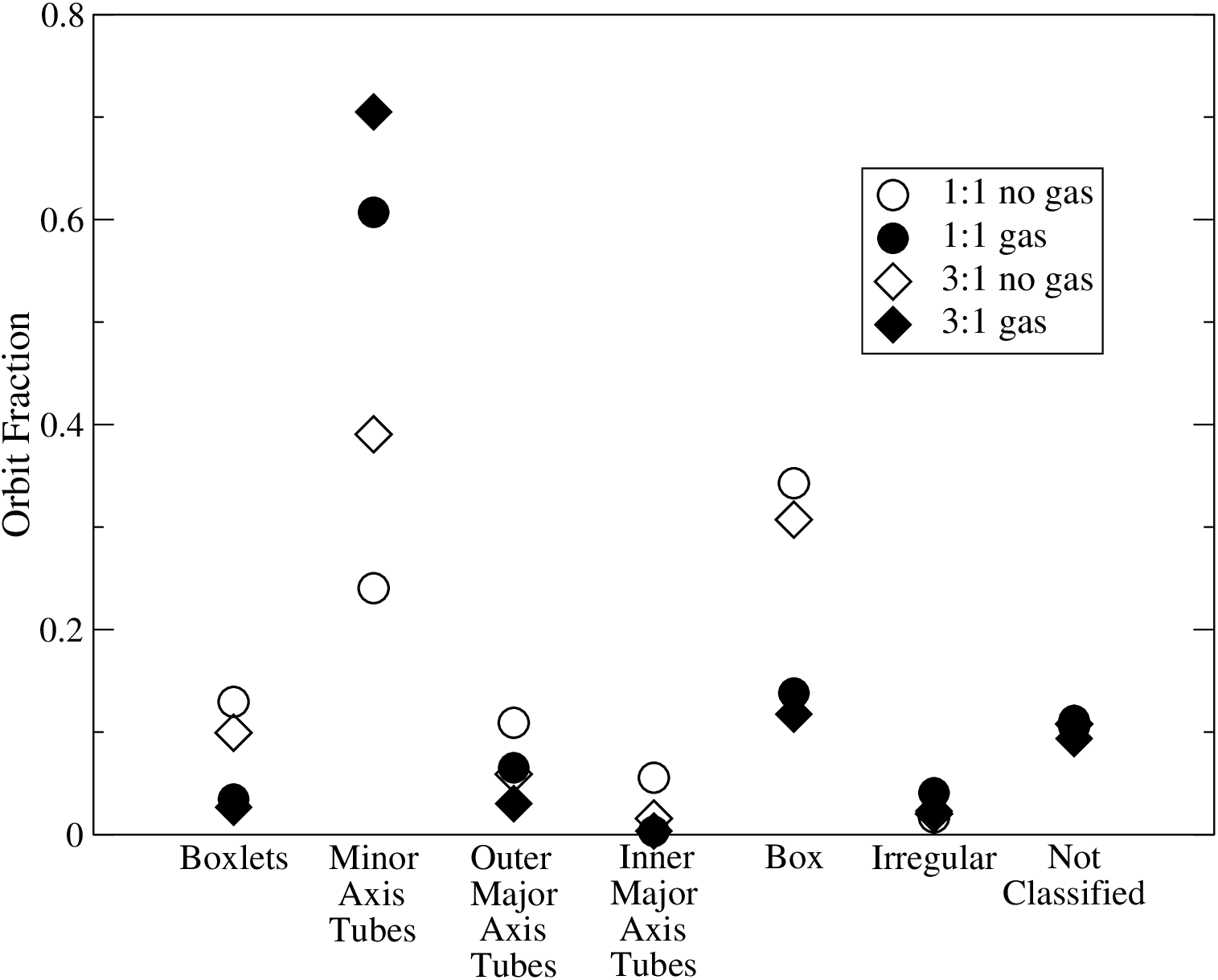, width=0.4\textwidth}
  \caption{Mean orbit fractions for the 40\% most tightly bound particles 
for all 1:1 (circles) and 3:1 (diamonds) 
remnants without (open symbols) and with gas (filled
  symbols).\label{mean_orbit}}
\end{center}
\end{figure}

\begin{figure}
\begin{center}
  \epsfig{file=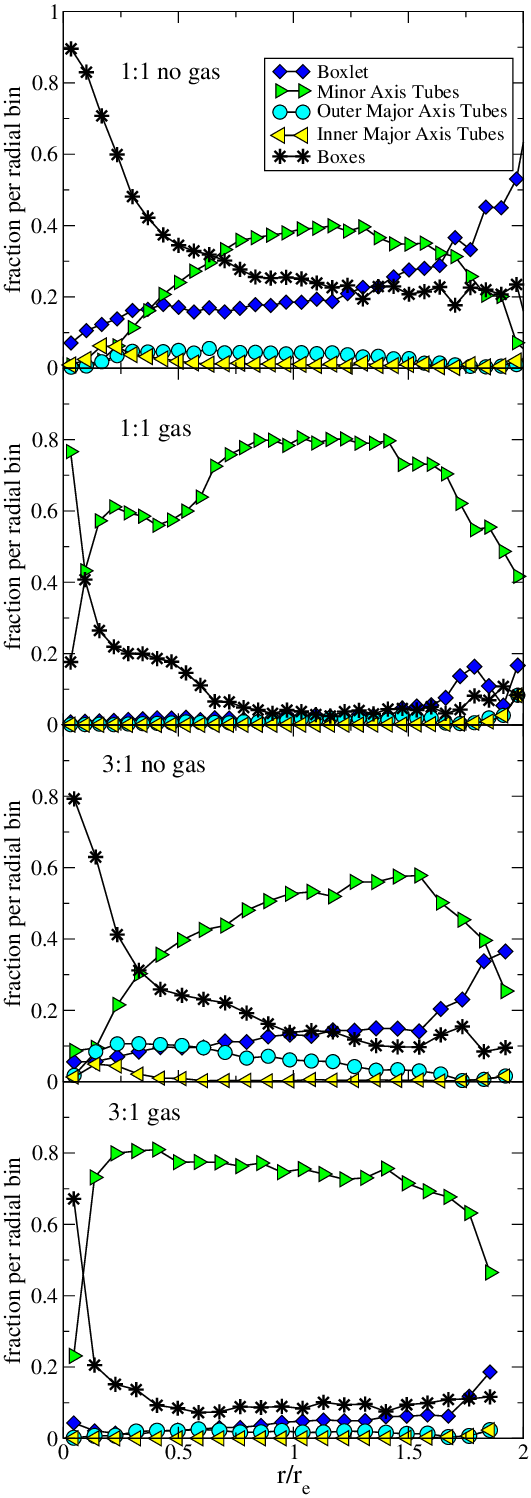, width=0.4\textwidth}
  \caption{Fraction of particles on different orbits in radial bins in units of
  $r_e$ for a typical 1:1 remnant without and
  with gas (two upper panels) and a typical 3:1 remnant with
  and without gas (two lower panels). Remnants with gas are more dominated by tube
  orbits. \label{orbitfraction_vs_radius}}  
\end{center}
\end{figure}

The classification of orbits follows the procedure presented
in JNB05 (and references therein) and is based on the method developed by
\citet{1998MNRAS.298....1C}. We will repeat only the most important
steps. After the merger remnant had settled into equilibrium we froze
the particle distribution and computed the potential using the 
Self Consistent Field (SCF) method \citep{1992ApJ...386..375H}. As initial
condition for the integration of the orbits we took the positions and
velocities of the particles at the final snapshot of the
simulation. With this method all orbit classes which exist in general
triaxial potentials can be identified. The orbits were
classified as minor-axis tubes, outer major-axis tubes, inner
major-axis tubes, boxes and boxlets. About 5\% -10\% of all orbits in 
every remnant could not be classified with this procedure. 

Every orbit class has distinct 
kinematical properties and shapes. Minor-axis tubes are the backbone
of oblate and disc-like systems and have a non-vanishing angular
momentum around the short axis of the potential. Major-axis tubes are
found in prolate or nearly spherical systems with a non-vanishing
angular momentum around the long axis. Boxes and boxlets have no mean
angular momentum and can be found close to the center of the
potential. 

We compared the population of each orbit class averaged over all
1:1 and 3:1 mergers, respectively, for all merger remnants with and
without gas. The result is shown in
Fig. \ref{mean_orbit}.  In both, 1:1 and 3:1 
mergers, gas drastically reduces the fraction of box orbits while the
fraction of minor-axis tubes increases by a factor of 2-3 (the slightly larger
central gas fractions of the simulations with higher halo resolution discussed in the 
previous section results in an increase of the tube orbit fraction of 3-5\%). This
confirms the observed trends towards an axisymmetric shape.
Major-axis tubes are also de-populated which is consistent with the
reduction of the prolateness. Box orbits are still found at
the smallest radii, but minor-axis tubes in general start to be the dominant
orbit class at smaller radii than in the dissipationless mergers
(Fig.\ref{orbitfraction_vs_radius}). This finding is in agreement with 
\citet{1996ApJ...471..115B} and \citet{1998giis.conf..275B}. 

The superposition of all stars, 
which are on different orbits, determine the projected
photometric and kinematic properties of the remnants which are
investigated in the following sections.

\section{Global photometric and kinematic properties}
\label{PHOTO}

\begin{figure*}
\begin{center}
  \epsfig{file=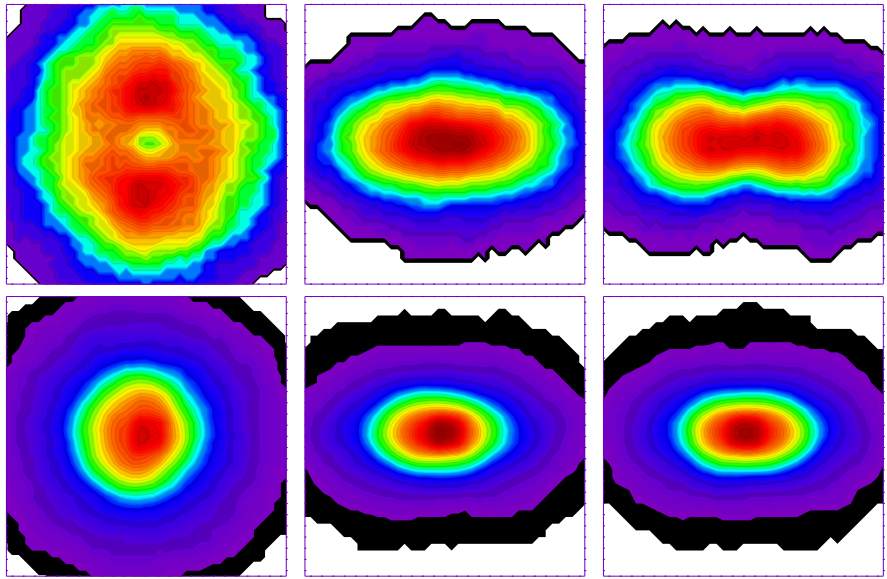, width=0.7\textwidth}
  \caption{Typical projected surface density of stellar particles on 
minor-axis tube orbits for a triaxial collisionless 1:1 remnant (upper row) and 
its more oblate counterpart with gas (bottom row). From left to right 
we show projections along the short, intermediate, and long axis.  The box 
length is $\approx 2 r_{\mathrm{eff}}$. Tube orbits, which dominate
around the effective radius, support a boxy/peanut isophotal shape in the triaxial 
potential of the collisionless remnant. In the more axisymmetric potential 
of the remnant with gas the shape is less boxy or even discy.    
\label{DENS_11MCS_5_ZT}}
\end{center}
\end{figure*}

We have performed the isophotal and kinematic analysis of 500 random
projections of every remnant following the procedure presented in
NB03. An artificial image of each projected 
remnant was created by binning the stars within the central 10 length 
units into $128 \times 128$ pixels,
smoothed with a Gaussian filter of standard deviation 1.5
pixels. The isophotes and their deviations from perfect ellipses were
then determined using a data reduction package kindly provided by Ralf
Bender. In this section the analysis was only performed for the
stellar particles. 
In the following we refer to the stellar particle distribution as
collisionless/without gas or dissipative/with gas depending on whether 
gas was present or not. 

\begin{figure}
\begin{center}
  \epsfig{file=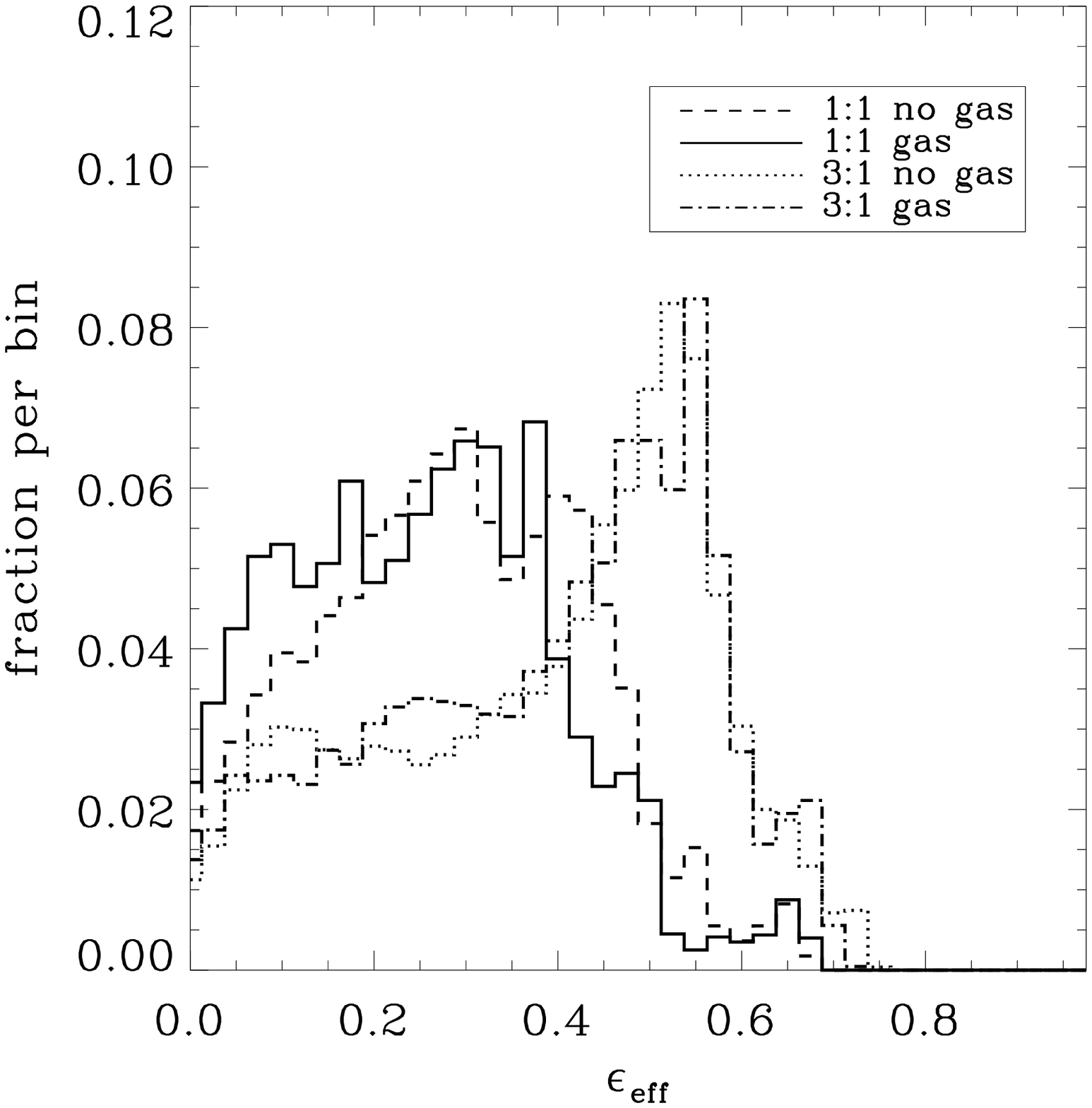, width=0.5\textwidth}
  \caption{Distribution of the projected effective ellipticities,
  $\epsilon_{\mathrm{eff}}$ for the stellar body of 1:1 remnants with (solid) and without
  (dashed) gas and for 3:1 remnants with (dashed-dotted) and without
  (dotted) gas.{\label{ellhist_pub_MGS}}}
\end{center}
\end{figure}

The characteristic ellipticity $\epsilon_{\mathrm{eff}}$ for
each projection is defined as the isophotal ellipticity  at
$1.5r_e$.  In Fig. \ref{ellhist_pub_MGS} we show the normalized
histograms of the ellipticity distribution for the stellar 1:1 and 3:1
remnants with and without gas. The distributions look similar for 3:1
remnants, whereas 1:1 remnants with gas have more projections with 
small ellipticities as expected if the main stellar body is more axisymmetric.   
Still, the distributions peak at similar ellipticities as mainly the ratio
of the intermediate to the long axis, $b/a$, is affected by the presence of 
gas (see Section \ref{SHAPES}).  

\begin{figure}
\begin{center}
  \epsfig{file=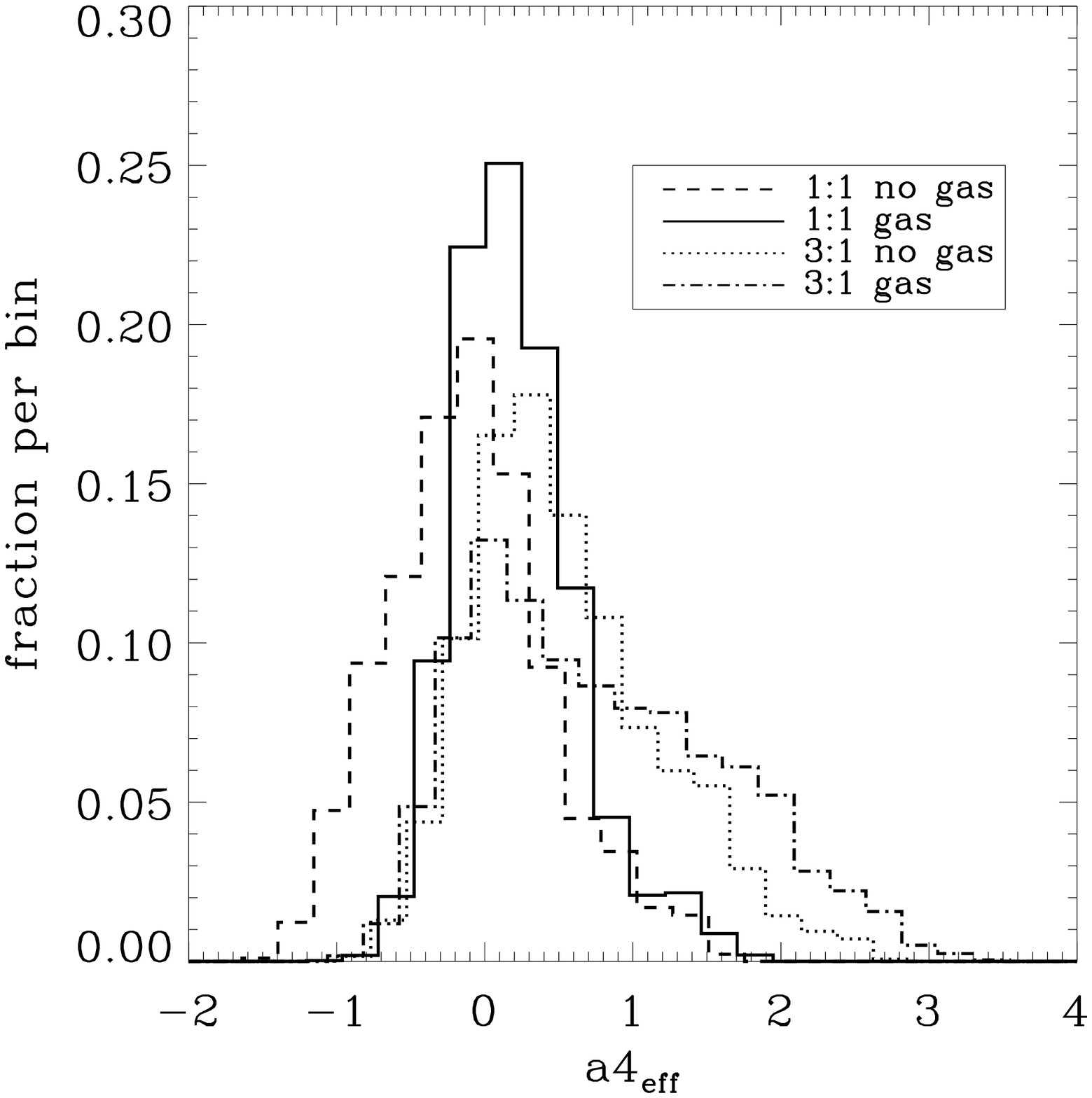, width=0.5\textwidth}
  \caption{Distribution of the projected effective isophotal shape,
  $a4_{\mathrm{eff}}$, of the stellar body for 1:1 remnants with (solid) and without
  (dashed) gas and for 3:1 remnants with (dashed-dotted) and without
  (dotted) gas.
\label{a4hist_pub_MGS}}
\end{center}
\end{figure}

\begin{figure*}
\begin{center}
  \epsfig{file=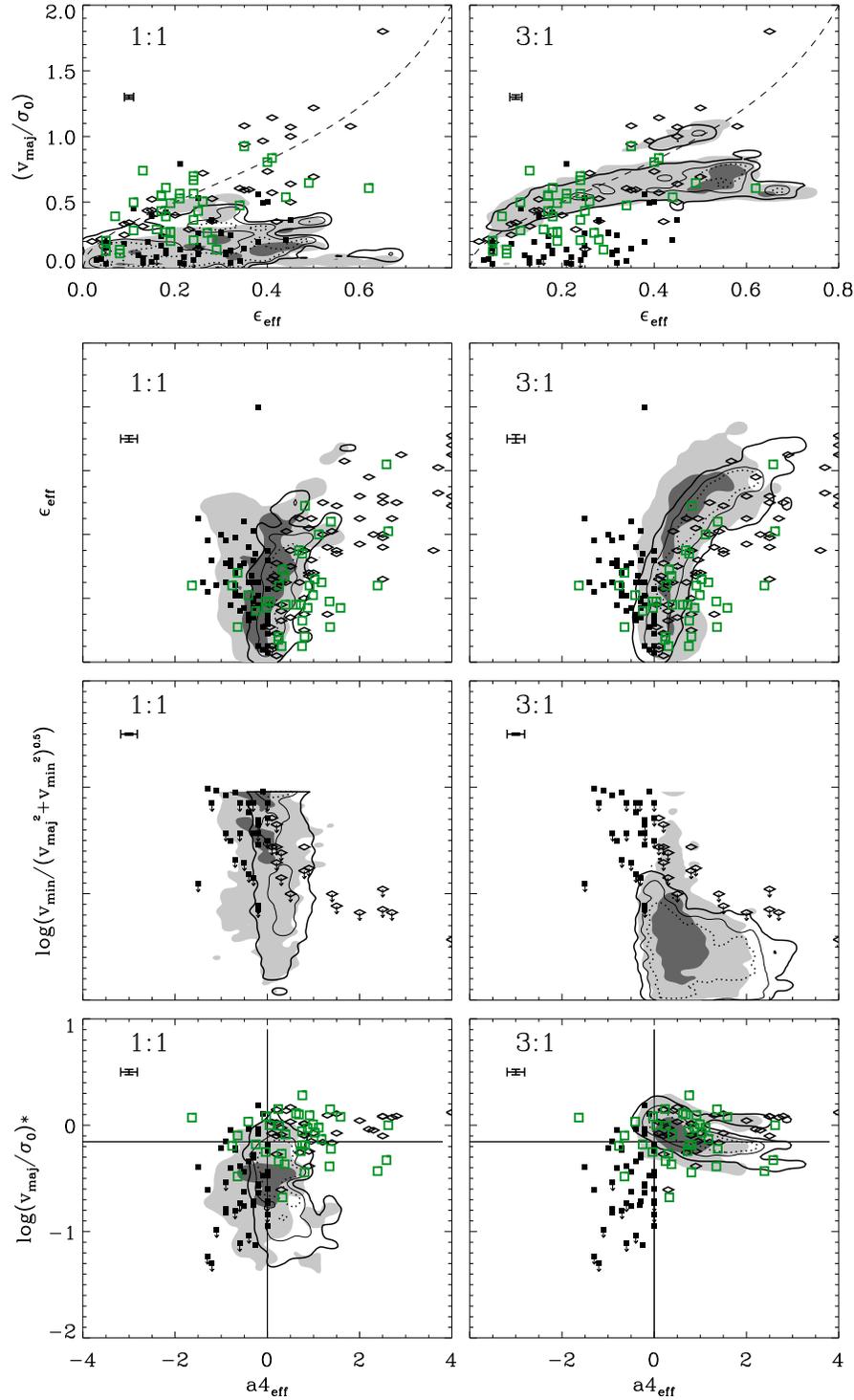, width=0.7\textwidth}
  \caption{Global photometric and kinematic properties of the stellar
    components of 1:1 and 3:1 merger remnants with and without
    gas. The contours indicate areas of 50\% (dotted line), 70\% (thin
    line) and 90\% (thick line) probability to detect a merger remnant
    with gas with the given properties. For comparison, the shaded
    areas indicate the 50\% (dark grey) and 90\% (bright grey)
    probabilities for collisionless remnants. 
    Data for observed boxy (filled
    squares) and discy (open diamonds) ellipticals have been kindly
    provided by Ralf Bender. Data for local merger remnants (open green squares) 
    are taken from \citet{2006astro.ph..4493R}. From top to bottom we show: ellipticity,
    $\epsilon_{\mathrm{eff}}$,  versus ${v_{\mathrm{maj}}/\sigma_0}$ (the theoretical value
    for a spheroid flattened by rotation is shown by the dashed line);
    effective isophotal shape, $a4_{\mathrm{eff}}$, versus
    $\epsilon_{\mathrm{eff}}$, versus the amount of minor-axis
    rotation, $v_{min} / \sqrt{v_{\mathrm{maj}}^2 + v_{min}^2}$, and
    versus the anisotropy
    parameter,$(v_{\mathrm{maj}}/\sigma_0)^*$. The stellar components 
    of remnants with gas (in particular 1:1) do not show boxy 
    isophotes any more and the 3:1
    remnants with gas are more discy and show less minor-axis
    rotation.  \label{overlay_all_MGS_pub}} 
\end{center}
\end{figure*}

The effective $a_4$-coefficient, $a4_{\mathrm{eff}}$, was computed as
the mean value of $a_4$ between $0.25 r_e$ and $1.0 r_e$, with $r_e$
being the projected spherical half-mass radius. In contrast to
NB03 we do not use the maximum value of $a_4$ in
case of a peaked distribution which did, however, not change the
results. The normalized histograms of the $a4_{\mathrm{eff}}$
distribution for all 1:1 and 3:1 remnants with and without gas is
shown in Fig. \ref{a4hist_pub_MGS}. In contrast to the collisionless case 
the 1:1 remnants with gas do not shown a significant number
of boxy projections any more. Their deviations from perfect
ellipses has decreased. For 3:1 remnants the effect of the gas is
weaker. There are now more projections which are significantly discy,
with $a4_{\mathrm{eff}}  > 2$.

The most important physical reason for the lack of boxy projections for mergers 
with gas is the different behaviour of minor-axis tube orbits in 
axisymmetric and triaxial potentials. As we have shown in Section 
\ref{SHAPES}, collisionless 1:1 remnants are more triaxial 
whereas stellar 1:1 remnants with gas are more axisymmetric. Minor-axis 
tube orbits in triaxial potentials can support a boxy/peanut 
isophotal shape in the projection along the minor axis and, more importantly, 
along the long axis (see Fig. 11 in JNB05). Minor-axis tubes are the dominant
orbit family around the effective radius of the collisionless remnants
(see Fig. \ref{orbitfraction_vs_radius}) and therefore are largely
responsible for the over all isophotal shape (see NB03 and JNB05). 
In the more axisymmetric potentials of the remnants with gas the 
minor-axis tubes look more elliptical/less boxy or even discy 
in all projections. This behaviour is qualitatively demonstrated
in Fig. \ref{DENS_11MCS_5_ZT} for a typical 1:1 remnant. In addition,
the fraction of box orbits and boxlets, which can support a boxy shape
at the center of collisionless remnants (see Fig. 11 in JNB05) is
significancy reduced (Fig. \ref{orbitfraction_vs_radius}). 
  
The central velocity dispersion $\sigma_0$ of every remnant was
determined as the average projected velocity dispersion of the
luminous particles inside a projected galactocentric distance of $
0.2r_e $ and we defined the characteristic rotational velocity along
the major axis, $v_{\mathrm{maj}}$, and the minor-axis,
$v_{\mathrm{min}}$, as the rotational velocity at $1.5r_e$ and $0.5
r_e$, respectively. The amount of minor-axis rotation was parameterized
as $v_{min} / \sqrt{v_{\mathrm{maj}}^2 + v_{min}^2}$
\citep{1985MNRAS.212..767B}. Minor-axis rotation in elliptical
galaxies, in addition to isophotal twist, has been suggested as a sign
for a triaxial shape of the main body of elliptical galaxies
\citep{1988A&A...195L...5W,1991ApJ...383..112F}. 
\begin{figure}
\begin{center}
  \epsfig{file=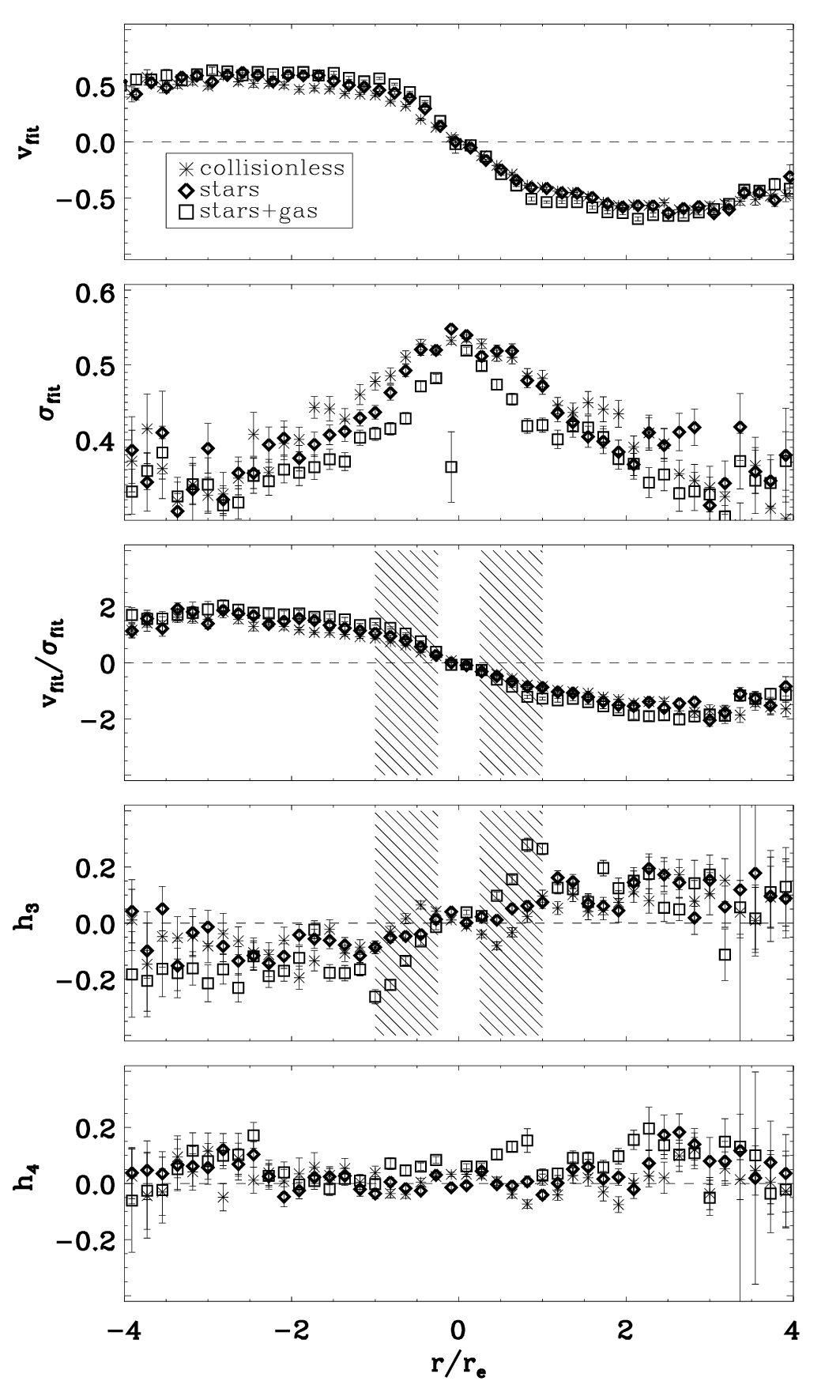, width=0.5\textwidth}
  \caption{Kinematical properties of a typical 3:1 merger remnant 
(collisionless, stars, stars+gas). The shaded area indicates the radial 
range where the local correlations between $h_3$ and 
$v_{\mathrm{fit}}/\sigma_{\mathrm{fit}}$ have been measured 
(see Fig. \ref{h3vs_local_comp_pub}).  If gas is present during the 
merger, $h_3$ changes sign. \label{losvd_comp_pub}}
\end{center}
\end{figure}

\begin{figure}
\begin{center}
  \epsfig{file=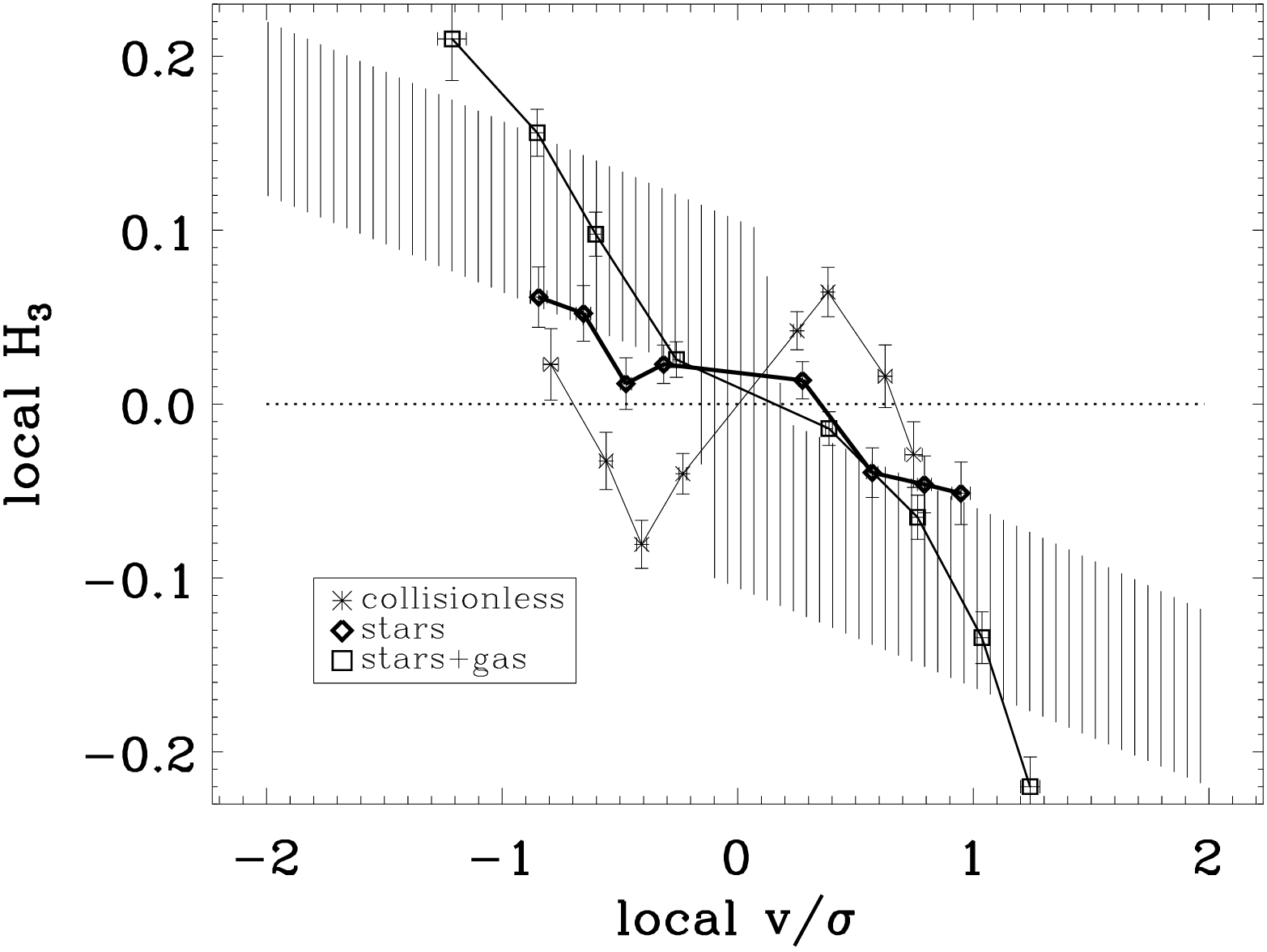, width=0.5\textwidth}
  \caption{Local correlation between $h_3$ and $v/\sigma$ for the same
 3:1 remnants as in Fig. \ref{losvd_comp_pub}. The collisionless
 remnant shows an anti-correlation between for low $v/\sigma$ and a
 correlation for the largest values. For the stellar remnant of the merger
 with gas $h_3$ and $v/\sigma$ is anti-correlated. Including the gas in
 the analysis leads to a stronger anti-correlation in good agreement
 with observations. The range of the observational data is indicated
 by the shaded area. \label{h3vs_local_comp_pub}}
\end{center}
\end{figure}

In Fig. \ref{overlay_all_MGS_pub} we summarize the properties of the stellar
components of the 1:1 and 3:1 merger remnants with gas in comparison
to the remnants without gas. The remaining gas is not included in this 
analysis. In the upper row we show the location of the remnants in the
${v_{\mathrm{maj}}/\sigma_0}-\epsilon_{\mathrm{eff}}$ plane. There
are no fundamental differences between remnants with and without
gas. The dichotomy found by \citet{1998giis.conf..275B}, \citet{1999ApJ...523L.133N} and 
NB03 that 1:1 mergers produce slowly rotating
remnants and 3:1 remnants produce fast rotating remnants is not changed by the 
presence of gas. In general the properties of 1:1 and 3:1 remnants 
are in very good agreement with recent observations of (predominantly
discy) nearby merger remnants by \citet{2006astro.ph..4493R} and discy
and boxy elliptical galaxies.

The distribution of the remnants in the
$\epsilon_{\mathrm{eff}}-a_4$ plane is shown in the
second row. As shown above the stellar components of 1:1 remnants with gas do not
appear boxy any more and can not explain the formation of boxy
ellipticals. For individual simulated mergers this trend was already
found by \citet{1997ApJ...478L..17B} and thereafter rediscovered by
\citet{2000MNRAS.312..859S}. Without assuming additional physics for
star formation all formerly boxy remnants are now close to elliptical
(see Fig. \ref{a4hist_pub_MGS}). The agreement of 3:1 remnants with nearby merger
remnants and discy elliptical galaxies is significantly better. 

In the third row of Fig. \ref{overlay_all_MGS_pub} we compare the
amount of minor-axis rotation as a function of the isophotal shape of
the collisionless remnants and the remnants with gas. There is a weak
trend for remnants with gas to show less minor-axis rotation than the
collisionless counterparts. This finding 
can be understood as the fraction of major-axis tubes, which are
mainly responsible for minor-axis rotation, is on average 
reduced due to the influence of gas (see Fig. \ref{mean_orbit}). 

The last row of Fig. \ref{overlay_all_MGS_pub} shows the location of the 
remnants in the $(v_{\mathrm{maj}}/\sigma_0)^*-a_4$ plane. Here
$(v_{\mathrm{maj}}/\sigma_0)^*$ is the traditional anisotropy
parameter (\citealp{1978MNRAS.183..501B},  see
\citet{2005MNRAS.363..937B,2005MNRAS.363..597B} for a revised version
of the anisotropy parameter and its application to N-body
simulations). On average the 1:1 remnants with gas have slightly 
larger values of $(v_{\mathrm{maj}}/\sigma_0)^*$ . Boxy and 
anisotropic remnants which we found for collisionless remnants did not form. 
In addition, we find more projected gas-remnants with small
$(v_{\mathrm{maj}}/\sigma_0)^*$ but discy isophotes than for the
pure collisionless remnants. In this regime no observed elliptical galaxies 
\citep{2006astro.ph..4493R} can be found. 
However, there are some S0 galaxies with similar 
properties \citep{1992ApJ...400L...5R}. A possible connection 
will be investigated in a subsequent paper. 3:1 remnants with gas have
similar properties to their collisionless counterparts. Remnants
of mergers with gas on average appear more discy and are in better agreement with 
observed discy ellipticals and nearby merger remnants.

\section{LOSVD analysis of the merger remnants}
\label{LOSVD}
\begin{figure}
\begin{center}
  \epsfig{file=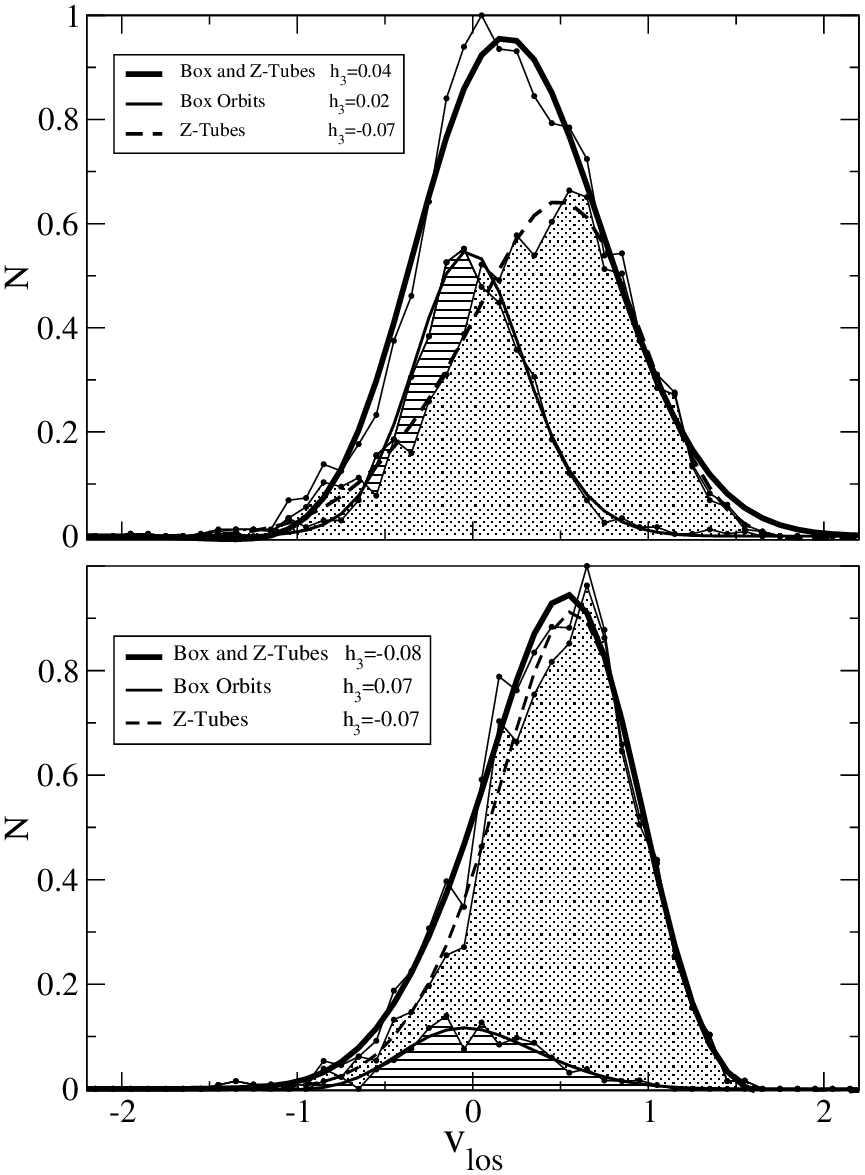, width=0.4\textwidth}
  \caption{Typical normalised velocity profiles at $\approx 0.75 r_{\mathrm{eff}}$ 
of the dominant population of Z-tubes and box orbits for a 
collisionless 3:1 remnant (upper panel) and the stellar component of its 
counterpart with gas (lower panel). Z-tubes have a negative $h_3$. In combination 
with the large fraction of box orbits of the collisionless remnant the overall 
LOSVD shows a broad leading wing. The remnant with gas is dominated by Z-tubes and 
has a steep leading wing, in agreement with observations of rotating elliptical galaxies.
 \label{vp}}
\end{center}
\end{figure}

To measure the LOSVDs of a merger remnant
we shifted the densest region of every two-dimensional projection to
the origin and placed a slit with a width of 0.4 unit lengths
along the apparent long axis of each projected remnant. 
The slit was then subdivided into grid cells of 0.15 unit
lengths. Thereafter we binned all particles falling within each cell in velocity along
the line-of-sight. The width of the velocity bins was set to
a value of 0.2 for line-of-sight velocities $v_{\mathrm{los}}$ in the
range $ -4 \le v_{\mathrm{los}} \le 4$. This resulted in 40 velocity
bins over the whole velocity interval. Using the binned velocity data
we constructed line-of-sight velocity profiles for each bin
along the slit. Subsequently we parameterized deviations from the
Gaussian shape using Gauss-Hermite basis
functions \citep{1993MNRAS.265..213G,1993ApJ...407..525V}.  
The kinematic parameters of each profile ($\sigma_{\mathrm{fit}}$, 
$v_{\mathrm{fit}}$, $h_3$, $h_4$) were then 
determined by least squares fitting (see \citealp{2001ApJ...554..291C}). 
The large number of simulated stellar particles ($>$ 100000) guaranteed that 
at least 1000 particles fall within each bin inside the effective radius.
We have analysed three sets of particle distributions: all star
particles of the collisionless simulation, the star particles of the
simulation with gas and, assuming that all gas particles have
transformed into stars, stars and gas particles of the simulation with
gas. In the following we refer to these sets of particles as:
collisionless, stars, and stars+gas.

\begin{figure*}
\begin{center}
  \epsfig{file=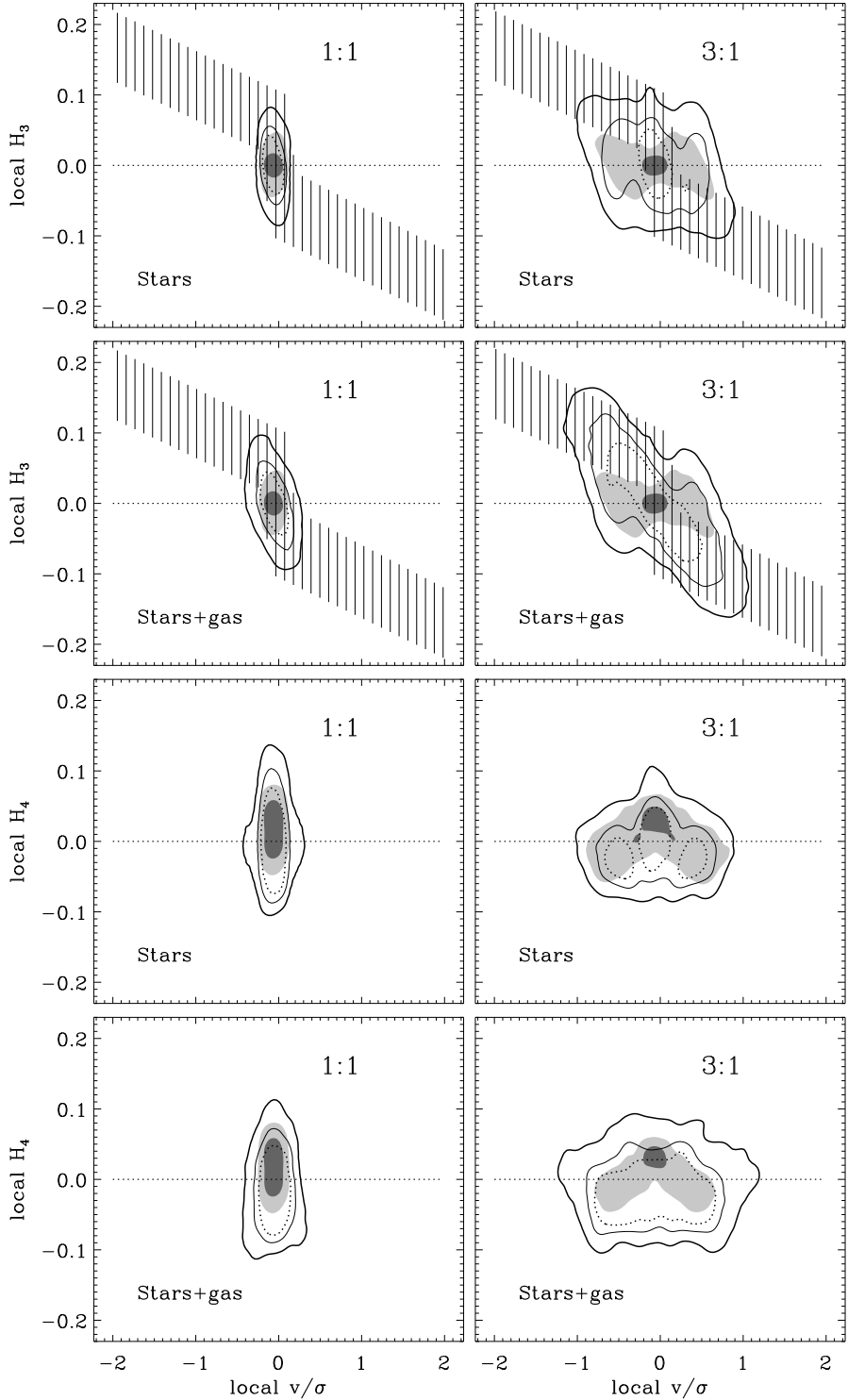, width=0.7\textwidth}
  \caption{{\it Upper row}: Local correlation between $h_3$ and $v/\sigma$
    between $0.25$ and $1r_e$ along the major axis for 1:1 and 3:1 merger
    remnants with and without 
    gas. The contours are as in Fig. \ref{overlay_all_MGS_pub}. The vertical lines indicate 
    the location of observational data by 
    \citet{1994MNRAS.269..785B}. The collisionless 1:1 remnants (left
    panel) rotate very slowly. The asymmetry of the LOSVD of
    the stars ($h_3$) increases if gas is added to the simulations.   
    Collisionless 3:1 remnants do not show the observed
    anti-correlation (right panel, shaded area). The corresponding
    stellar components of the mergers with gas (contours) have 
    larger maximum values
    of $v/\sigma$ and show stronger asymmetries with  a clear
    indication for a change in the tilt. {\it Second row}: Including
    the gas (assuming it has transformed into stars after the merger
    was complete) the distribution for equal mass remnants becomes weakly
    tilted (left panel) whereas the distribution for 3:1 remnants
    (right panel) shows a clear tilt and extends to a local $v/\sigma
    \approx 1$ in good agreement with observations. {\it Third row}:
    Same as above but for $h_4$. Most collisionless remnants (shaded
    areas) have a positive local values of $h_4$ whereas the remnants
    with gas show equal amount of positive and negative values
    (contours). For 3:1 remnants there is a trend for regions with
    higher $v/\sigma$ to have negative $h_4$. {\it Bottom panel}:
    Adding gas to the analysis results in an over-all shift to more
    negative values of $h_4$.  \label{overlay_h34vs_all_pub}}
\end{center}
\end{figure*}

As a prototypical example Fig. \ref{losvd_comp_pub} shows the 
observables $v_{\mathrm{fit}}$, 
$\sigma_{\mathrm{fit}}$, $v_{\mathrm{fit}}/\sigma_{\mathrm{fit}}$, 
$h_3$, and $h_4$ together with the bootstrapping errors as a function
of radius for a 3:1 merger remnant. We compare the kinematics 
of the collisionless remnant with the stars and stars+gas of the simulated 
remnant with gas. The line-of-sight velocity is only weakly affected by 
the presence of gas. There is a slightly steeper gradient at the center 
for the merger with gas. The velocity dispersion profile (stars+gas) shows a clear signature 
of the extended gas disc: it decreases faster with increasing
radius. At the center where the gas has formed a cold dense blob the
differences in the dispersion are most significant. 

The most interesting feature can be seen for
$h_3$ which characterises the asymmetry of the LOSVD. In the collisionless case 
$h_3$ and $v_{\mathrm{fit}}$ have the same sign - and nearly the same slope -
inside half an effective radius. Towards one effective radius the slope 
for $h_3$ changes and at larger radii $v_{\mathrm{fit}}$ and $h_3$ are 
anti-correlated (Fig. \ref{losvd_comp_pub}). This behaviour
is typical for all 3:1 remnants \citep{2001ApJ...555L..91N,2000MNRAS.316..315B} 
and it is in contradiction to observed elliptical galaxies \citep{1994MNRAS.269..785B} where 
$h_3$ and $v_{\mathrm{fit}}/\sigma_{\mathrm{fit}}$ are anti-correlated inside the 
effective radius (indicated by the shaded area in
Fig. \ref{h3vs_local_comp_pub}).  The slope of the 
correlation between $h_3$ and $v_{\mathrm{fit}}/\sigma_{\mathrm{fit}}$
changes if gas is present during the merger. In the
region of interest $h_3$ and $v_{\mathrm{fit}}/\sigma_{\mathrm{fit}}$ are now 
weakly anti-correlated for the stars and strongly anti-correlated for stars+gas 
(see Fig. \ref{h3vs_local_comp_pub}). In Fig. \ref{vp} we show
the velocity distribution and the Gauss-Hermite fits for box orbits and Z-tubes, 
which dominate the remnants. The collisionless 3:1 remnant is compared to its 
counterpart with gas (same initial orientation). For both remnants the 
Z-tubes have a negative $h_3$ which corresponds to a steep leading wing. In combination 
with the large fraction of box orbits the resulting LOSVD of the 
collisionless remnant has a broad leading wing. \citet{2000MNRAS.316..315B} have shown a 
very similar distribution for their remnants. The stellar component of the remnant 
with gas has a much lower fraction of box orbits and the resulting LOSVD is dominated  
by the steep leading wing of the minor-axis tubes. 

The fourth order coefficient, $h_4$, characterizes a LOSVD  that is more
peaked than a Gaussian if positive and less peaked if negative 
(Fig. \ref{losvd_comp_pub}). For the 3:1 remnant chosen here $h_4$ appears 
to be positive at the center and negative around the effective radius, a feature that 
is less prominent for simulations with gas. In this paper 
we show the results for $h_4$ for completeness, however we will not discuss it in 
greater detail.

In Fig. \ref{overlay_h34vs_all_pub} we have summarised the local correlations
between $h_3$ and $v/\sigma$ in the range of 0.25 to  1 $r_e$ along the projected 
major axis for 500 random projections for all 1:1 and 3:1 merger
remnants with and without gas. 
The collisionless 1:1 remnants (left panel) rotate very slowly and are consistent 
with observations. The asymmetry of the LOSVD of the stars ($h_3$) 
increases if gas is added to the simulations. Rotating collisionless 3:1 remnants do 
not show the observed anti-correlation (right panel, shaded area). The corresponding
stellar components of the mergers with gas (contours) have larger maximum values of 
$v/\sigma$ and show stronger asymmetries with  a clear indication for a change
of tilt. Including the gas (assuming it has transformed 
into stars after the merger was complete), the distribution for equal-mass remnants 
becomes weakly tilted whereas the distribution for 3:1 remnants shows a clear 
tilt and extends to a local $v/\sigma
\approx 1$ in good agreement with observations. 

The lower panels of Figure \ref{overlay_h34vs_all_pub} show the correlation with
$h_4$. Most collisionless remnants have positive local values of $h_4$ whereas the remnants
with gas show equal amount of positive and negative values. For 3:1 collisionless
remnants there is 
a trend for regions with higher $v/\sigma$ to have negative $h_4$. Adding gas to the 
analysis results in an over-all shift to more negative values of $h_4$.

\section{Summary and Discussion} 
\label{SUMMARY}
We have presented a statistical sample of simulations of disc galaxy mergers
with a 10\% fraction of gas in the progenitor discs. The properties 
of the merger remnants have been compared to the properties of a similar set 
of simulations without gas. The effect of star-formation was not included. 
The presence of a dissipative component changes  
the shape and the orbital content of the stellar component of the merger remnants. 
The fraction of box and boxlet orbits which dominate the inner parts of collisionless 
remnants is significantly reduced if gas is included. The fraction of outer major-axis 
tubes is reduced as well and inner major-axis tubes disappear
completely. In remnants with gas most stars move on minor-axis tube
orbits. The change of orbits is caused by gas that settles at the center of the remnants 
where it deepens the potential well at the same time making it more axisymmetric. 
In this environment box orbits cannot exist as they are only supported in 
triaxial potentials. We find that the intermediate axis of the stellar distribution 
is most strongly affected and the stellar remnants with gas are more 
axisymmetric. These results are in good qualitative agreement with the findings of 
\citet{1996ApJ...471..115B} and \citet{1998giis.conf..275B}. However, with respect to 
resolution, statistical completeness and comparison to observations our study goes beyond
previous investigations.   

The isophotal shape of equal-mass remnants is strongly affected by gas. 
Around the effective radius minor-axis tubes - as well as box-orbits and boxlets at 
smaller radii - in the triaxial potential of collisionless remnants 
support boxy isophotal shapes. In the remnants with gas 
the fraction of box orbits is reduced and the now dominant tube orbits 
appear more elliptical or even discy. 
Statistically, equal-mass mergers with gas do not produce boxy remnants, but they still have
a small anisotropy parameter which is in conflict with observations of ellipticals. 
This is only valid in the limiting case of no starformation during the merger. 
Realistically, equal-mass disc mergers like nearby ULIRGs do experience bursts of 
star formation \citep{1998ApJ...498..579G,2001ApJ...563..527G}. It is, however, still 
unclear how much of the available gas is transformed into stars at which stage of the merger. 
For individual equal-mass merger simulations it has been shown by \citet{1997ApJ...478L..17B} 
and \citet{2000MNRAS.312..859S} that boxy remnants can be formed if the star-formation 
efficiencies were chosen to be high leading to an evolution that is similar to 
the collisionless case \citep{1999ApJ...523L.133N}. Disky remnants form for low 
efficiencies similar to the results with gas presented here. 3:1 remnants with gas 
are slightly more discy (see e.g. \citealp{1998ApJ...502L.133B} for an unequal-mass merger 
simulation including star formation) but in 
general resemble their collisionless counterpart and are in good agreement with observations
of discy elliptical galaxies \citep{1988A&AS...74..385B,2006astro.ph..5319H} as well as 
nearby merger remnants \citep{2006astro.ph..4493R}. We expect that the effect of star 
formation on the isophotal shape of unequal mass remnants is much weaker 
(see e.g. \citealp{2005A&A...437...69B}).  

The shape of the LOSVD of the stars is significantly influenced by 
the presence of gas. Collisionless remnants have LOSVDs which are close to 
Gaussian or have broad leading wings. In contrast the stellar component of mergers 
with gas shows are clear tendency for steep leading wings. 
For remnants with a significant amount of rotation (e.g. 3:1 remnants) this results in 
an anti-correlation between $h_3$ and $v/\sigma$, especially if the gas component that 
has formed extended discs is included in the analysis after the merger is complete.  
This is in good agreement with observations of discy,   
fast rotating ellipticals \citep{1994MNRAS.269..785B} and might indicate low star-formation 
efficiencies during the merger event itself. The LOSVDs  
of equal-mass merger remnants are also consistent with boxy, slowly rotating ellipticals.

If mergers of discs galaxies have indeed formed elliptical galaxies in the not to recent 
past our results confirm the scenario of \citet{1996ApJ...464L.119K} that gas
dissipation becomes more important the more discy the isophotes of early-type galaxies are 
and the faster they rotate. Some equal-mass merger remnants with gas still have a small 
anisotropy parameter and discy isophotes. Either this conflict with observations can be 
solved by efficient star formation during the merger or it indicates that massive,
boxy ellipticals formed in dissipationless mergers from predominantly stellar
progenitors which either were early-type disc galaxies or ellipticals themselves
\citep{1999ApJ...523L.133N,2003ApJ...597L.117K,2005MNRAS.359.1379K,
2005MNRAS.361.1043G,2006ApJ...636L..81N}. Recent high resolution direct 
numerical simulations of the formation of field elliptical galaxies from 
cosmological initial conditions indicate that early-type galaxies after they have 
formed at high redshift during a phase of intensive merging can thereafter grow 
by accretion of mainly stellar satellites or minor mergers \citep{2005astro.ph.12235N}.     

Combining the results from NB03 and the
present study we can conclude that mergers of typical disc galaxies with bulges
can have formed giant elliptical of low and intermediate mass in the past. 
The agreement of kinematic and photometric properties of 1:1 and in particular 
3:1 disc mergers with nearby merger remnants is striking. In combination with 
results from the simulations presented here \citet{2006ApJ...638..745D} have been 
able to show that ongoing disc mergers with ULIRG activity have mass ratios 
between 1:1 and 3:1. Furthermore the properties of the simulated remnants are in good 
agreement with the kinematics and isophotal shapes of nearby merger remnants 
\citep{2006astro.ph..4493R}. A particularly interesting result is that 
\citet{2006astro.ph..4493R} find merger remnants which are discy and 
anisotropic, a regime that is populated by simulated remnants but not by virialised old 
elliptical galaxies (see their paper for a detailed discussion). The direct comparison of real 
nearby mergers with merger simulations are a powerful tool to constrain theories 
on elliptical galaxy formation and the effect of star formation and black-hole 
formation \citep{2006astro.ph..7468D} in the local universe 
which then can be applied to simulations at all redshift ranges.  

In the present simulations the conversion of gas into stars has not been included and 
only progenitors with a small gas fraction have been considered
as otherwise neglecting star formation would hardly be justified. We find
important signatures of the small gas component in the merger remnants. 
The question of how stars form and how stellar energetic feedback or central
black hole heating affect the multi-phase interstellar medium is still poorly 
understood (for a summary see e.g. \citealp{2004ARA&A..42..211E}),
leading to large uncertainties in models of galaxy formation and evolution. 
Recent simulation e.g. by \citet{2005ApJ...620L..79S} and 
\citet{2005Natur.433..604D}
however demonstrate their importance in understanding the
formation of elliptical galaxies in greater details. The orbital content of 
the remnants will be influenced by the timing of black-hole formation and 
the efficiency of feedback which determines the amount of gas that can make it to the center
leading to the effects presented here. It will be interesting
in future work to explore the effect of these processes on the internal
orbital structure and projected properties of elliptical galaxies.
From a kinematic point of view disc merger remnants appear very similar to observed ellipticals, 
although questions regarding the age and metalicity of the stellar populations have to be 
addressed in the future. Analytical models of disc formation (e.g. \citealp{2006MNRAS.366..899N})
which contain all the information about stellar ages and metallicitites of the progenitor discs 
and semi-analytical modeling \citep{2005MNRAS.359.1379K,2006MNRAS.370..902K} in combination 
with detailed merger simulations \citep{2006ApJ...636L..81N} can be used to place further 
constraints on the disc merger hypothesis.

\section*{Acknowledgments}
This work was supported by the DFG priority program 1177. TN is grateful 
for interesting discussions with Kalliopi Dasyra, Gregory Novak and Barry Rotherg. 
He also thanks Barry Rotherg for kindly providing his observational data prior to 
publication. We thank the referee for valuable comments on the manuscript. 

\bibliographystyle{mn2e}
\bibliography{/data/e0/users/naab/PAPER/REFERENCES/references}

\label{lastpage}
\end{document}